\PassOptionsToPackage{unicode}{hyperref}
\PassOptionsToPackage{hyphens}{url}
\documentclass[
]{article}
\usepackage{xcolor}
\usepackage{amsmath,amssymb}
\usepackage{iftex}
\ifPDFTeX
  \usepackage[T1]{fontenc}
  \usepackage[utf8]{inputenc}
  \usepackage{textcomp} 
\else 
  \usepackage{unicode-math} 
  \defaultfontfeatures{Scale=MatchLowercase}
  \defaultfontfeatures[\rmfamily]{Ligatures=TeX,Scale=1}
\fi
\usepackage{lmodern}
\ifPDFTeX\else
\fi
\IfFileExists{upquote.sty}{\usepackage{upquote}}{}
\IfFileExists{microtype.sty}{
  \usepackage[]{microtype}
  \UseMicrotypeSet[protrusion]{basicmath} 
}{}
\makeatletter
\@ifundefined{KOMAClassName}{
  \IfFileExists{parskip.sty}{%
    \usepackage{parskip}
  }{
    \setlength{\parindent}{0pt}
    \setlength{\parskip}{6pt plus 2pt minus 1pt}}
}{
  \KOMAoptions{parskip=half}}
\makeatother
\usepackage{longtable,booktabs,array}
\usepackage{calc} 
\usepackage{etoolbox}
\makeatletter
\patchcmd\longtable{\par}{\if@noskipsec\mbox{}\fi\par}{}{}
\makeatother
\IfFileExists{footnotehyper.sty}{\usepackage{footnotehyper}}{\usepackage{footnote}}
\makesavenoteenv{longtable}
\usepackage{graphicx}
\makeatletter
\newsavebox\pandoc@box
\newcommand*\pandocbounded[1]{
  \sbox\pandoc@box{#1}%
  \Gscale@div\@tempa{\textheight}{\dimexpr\ht\pandoc@box+\dp\pandoc@box\relax}%
  \Gscale@div\@tempb{\linewidth}{\wd\pandoc@box}%
  \ifdim\@tempb\p@<\@tempa\p@\let\@tempa\@tempb\fi
  \ifdim\@tempa\p@<\p@\scalebox{\@tempa}{\usebox\pandoc@box}%
  \else\usebox{\pandoc@box}%
  \fi%
}
\def\fps@figure{htbp}
\makeatother
\usepackage{bookmark}
\IfFileExists{xurl.sty}{\usepackage{xurl}}{} 
\hypersetup{
  hidelinks,
  pdfcreator={LaTeX via pandoc}}

\author{}
\date{}

\begin{document}

\textbf{THE INTERSTELLAR SIGNATURE: A COMPUTATIONAL FRAMEWORK FOR OPEN
SOURCE INTERSTELLAR TRACKING}

Pancha Narayan Sahu , Independent Student Researcher, NEPAL\\
panchanarayansahu00@gmail.com

{\def\LTcaptype{none} 
\begin{longtable}[]{@{}
  >{\raggedright\arraybackslash}p{(\linewidth - 4\tabcolsep) * \real{0.1920}}
  >{\raggedright\arraybackslash}p{(\linewidth - 4\tabcolsep) * \real{0.0289}}
  >{\raggedright\arraybackslash}p{(\linewidth - 4\tabcolsep) * \real{0.7791}}@{}}
\toprule\noalign{}
\begin{minipage}[b]{\linewidth}\raggedright
Keywords
\end{minipage} & \begin{minipage}[b]{\linewidth}\raggedright
\end{minipage} & \begin{minipage}[b]{\linewidth}\raggedright
Abstract
\end{minipage} \\
\midrule\noalign{}
\endhead
\bottomrule\noalign{}
\endlastfoot
Interstellar Objects

3I/ATLAS

Orbital Mechanics

Simulation

SOLAR SYSTEM OBJECTS

Data Science

Astrophysics

Astronomy

Python & & Interstellar objects, such as 1I/`Oumuamua and 2I/Borisov,
offer a unique window into the formation and evolution of other star
systems, yet tracking and analyzing their trajectories remains largely
restricted to specialized institutions. Interstellar and solar system
datasets are often large, complex, and difficult to navigate, limiting
their usability for developers, researchers, and enthusiasts. To address
this, we present The Interstellar Signature: A Computational Framework
for Open-Source Interstellar Tracking, implemented through a web-based
platform.

Interstellar Signature serves as a bridge between raw, unstructured
astronomical data and an intuitive, developer-friendly interface. This
framework integrates live astronomical data from public repositories and
APIs with physics-based simulation techniques to model and visualize the
motion of both solar system and interstellar objects in real time. The
platform provides interactive visualizations, comparative analysis of
interstellar and solar system objects, and modular tools that allow
users to explore, modify, and extend the framework for their own
research purposes.

As an open-source project, it encourages experimentation, collaborative
development, and direct engagement with complex datasets that would
otherwise be difficult to interpret. Interstellar Signature is one of
the core projects under NexusCosmos, an open-source ecosystem envisioned
as a ``Linux for the space race,'' designed to democratize access to
space science data and interactive tools. By transforming inaccessible
datasets into a manipulable, visual, and educational experience,
Interstellar Signature empowers developers and researchers to explore
interstellar phenomena with clarity, flexibility, and creativity.

Future extensions will incorporate AI-driven modules for trajectory
prediction, anomaly detection, and enhanced visualization. By combining
open-source accessibility, computational rigor, and interactive
simulation, Interstellar Signature democratizes interstellar tracking,
making advanced space research available to a broader scientific and
educational community. This framework represents a step toward bridging
professional astronomical research and public engagement through
technology. \\
& & \\
\end{longtable}
}

\subsection{INTRODUCTION}\label{introduction}

The detection of interstellar objects (ISOs) has expanded our
understanding of cosmic dynamics, yet their analysis remains hindered by
the fragmented and complex nature of available astronomical data. While
public observatories and research agencies release extensive datasets,
their raw forms---often heterogeneous, unstructured, and difficult to
interpret---pose significant challenges for computational exploration.
This creates a gap between data availability and practical
accessibility, particularly for independent researchers and developers
seeking to examine ISO trajectories or perform comparative analyses with
solar system objects.

This research investigates methods for transforming irregular,
multi-source datasets into an interactive and coherent format, enabling
users to visualize, compare, and analyze orbital behaviors through
computational tools. The framework integrates live ephemeris and
observational data, parses them into standardized structures, and
renders the results within a real-time visualization engine designed for
exploratory research. The focus of this study is not on constructing new
physical models, but rather on enhancing the interpretability and
usability of existing astronomical data. By prioritizing data structure,
visualization, and interactivity, Interstellar Signature demonstrates
how open computational frameworks can lower barriers to interstellar
research.

As an open-source project, Interstellar Signature provides full
transparency and flexibility. Developers can extend its modules, build
upon existing visualizations, or integrate their own data sources. This
approach not only enhances accessibility but also fosters collaboration
among programmers, educators, and astronomy enthusiasts. The project was
developed through extensive research on existing platforms, aiming to
combine their strengths while addressing their limitations. It utilizes
live JPL Horizons (JPL, 2025 ) data for selected objects and simulates
them to create a 3D model of real space, allowing researchers,
developers, and students to visualize astronomical phenomena
interactively.

Additionally, users gain access to refined data and can compare
different interstellar objects. The processed data is available in JSON
format, ready for use in custom simulations. Independent researchers
like me can leverage it for comparative analysis or personal research.
The beauty of open source lies in its limitless potential---it empowers
anyone to modify, extend, and innovate freely.

\includegraphics[width=5.04931in,height=2.2214in,alt={C:\textbackslash Users\textbackslash hp\textbackslash OneDrive\textbackslash Documents\textbackslash reseach\_graphs\textbackslash overview.png}]{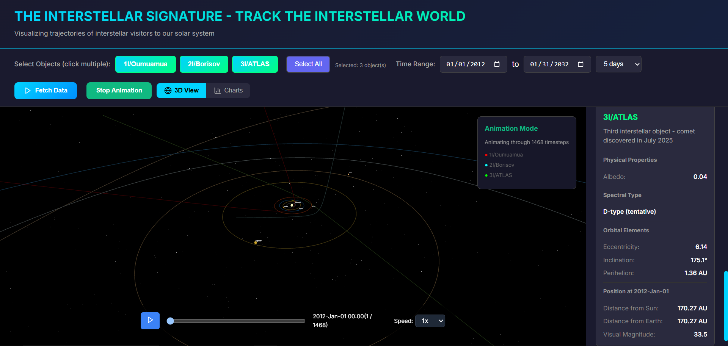}

\emph{Fig.1 (Shows the 3D-view feature of the application)}

\includegraphics[width=5.025in,height=2.40035in,alt={C:\textbackslash Users\textbackslash hp\textbackslash OneDrive\textbackslash Documents\textbackslash reseach\_graphs\textbackslash overview\_2.png}]{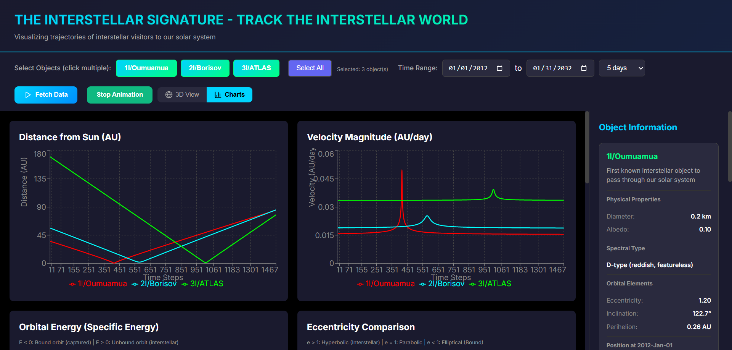}

\emph{Fig.2 (shows the graph view of the application)}

\subsection{BACKGROUND}\label{background}

Existing platforms such as NASA Eyes (NASA, NASA Eyes on the Solar
System, 2025), ESA's Gaia Archive Agency (ESA) (ESA, 2025), JPL Horizon
(JPL, 2025 ), Celestia (Team C. D., 2025) , and Space Engine (Team S. ,
2025) have advanced both public and academic engagement with
astronomical data. However, most of these systems are either
institutional tools built for scientific precision or visualization
platforms designed primarily for presentation. As a result, they often
lack flexibility, open access, or modularity for independent researchers
and developers.

While JPL provides highly accurate data, it can be complex to integrate
into custom workflows. Similarly, visualization tools like Celestia
(Team C. D., 2025) and Space Engine (Team S. , 2025) limit user
modification and interactivity with live datasets. This creates a
persistent gap between data availability and practical accessibility.

Interstellar Signature addresses this gap by introducing an open-source,
modular, and real-time visualization framework that merges scientific
data with interactive simulation---making interstellar research more
approachable, adaptable, and collaborative.

\subsection{}\label{section}

\subsection{METHODOLOGY}\label{methodology}

\subsubsection{Data Acquisition}\label{data-acquisition}

The trajectory simulation and analysis of the interstellar object relied
on two primary sources of data: the NASA JPL Horizons (JPL, 2025 )
system and the Planetary Data System (PDS) ((PDS), 2025). The JPL
Horizons system (JPL, 2025 ) provides high-precision ephemeris data,
including heliocentric positions, velocities, and orbital elements, for
small bodies within and beyond the Solar System. Ephemeris data were
obtained by querying the Horizons API (JPL, 2025 ) using the
object-specific identifier, a defined time range, step size, and
observer location. The API returned data in CSV format, which were
parsed to extract Cartesian coordinates and relevant orbital parameters.
This information formed the foundational dataset for numerical
integration and trajectory computation.

Complementary physical and discovery metadata were obtained from the
Planetary Data System ((PDS), 2025) JSON files from the PDS contain
information on object size, spectral properties, discovery
circumstances, and other relevant characteristics. These data were
merged with the JPL ephemeris using a defined merging protocol to create
a comprehensive dataset suitable for both computational analysis and
visualization. The integration process ensured consistency in units,
coordinate frames, and temporal references, enabling seamless use in
subsequent simulation and visualization steps.

The combined dataset, consisting of positional, orbital, and physical
properties, provides a complete basis for the study of the interstellar
object's dynamics, allowing for precise computation of trajectories,
velocities, and energy characteristics over the observational time span.

\subsubsection{Orbital Mechanics
Implementation}\label{orbital-mechanics-implementation}

The motion of the interstellar object was modeled under the classical
two-body approximation, treating the Sun as the central gravitational
body and the object as a test particle. The trajectory was defined using
standard orbital elements: semi-major axis (a), eccentricity (e),
inclination (i), longitude of ascending node (Ω), argument of perihelion
(ω), and mean anomaly (M).

The mean motion n of the object was computed from the orbital period `P'
as:

\includegraphics[width=0.75in,height=0.34722in,alt={\{"mathml":"\textless math style=\textbackslash"font-family:stix;font-size:16px;\textbackslash" xmlns=\textbackslash"http://www.w3.org/1998/Math/MathML\textbackslash"\textgreater\textless mstyle mathsize=\textbackslash"16px\textbackslash"\textgreater\textless mi\textgreater n\textless/mi\textgreater\textless mo\textgreater\&\#xA0;\textless/mo\textgreater\textless mo\textgreater=\textless/mo\textgreater\textless mo\textgreater\&\#xA0;\textless/mo\textgreater\textless mfrac\textgreater\textless msup\textgreater\textless mn\textgreater360\textless/mn\textgreater\textless mo\textgreater\&\#x2218;\textless/mo\textgreater\textless/msup\textgreater\textless mi\textgreater P\textless/mi\textgreater\textless/mfrac\textgreater\textless/mstyle\textgreater\textless/math\textgreater","truncated":false\}}]{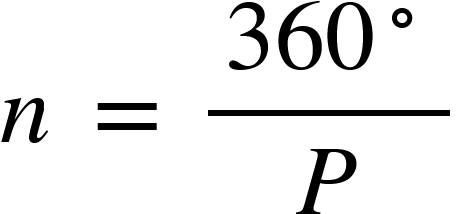}
(1)

and the mean anomaly at any time `t' was obtained using the relation:

\includegraphics[width=1.91667in,height=0.16667in,alt={\{"mathml":"\textless math style=\textbackslash"font-family:stix;font-size:16px;\textbackslash" xmlns=\textbackslash"http://www.w3.org/1998/Math/MathML\textbackslash"\textgreater\textless mstyle mathsize=\textbackslash"16px\textbackslash"\textgreater\textless mi\textgreater M\textless/mi\textgreater\textless mo\textgreater=\textless/mo\textgreater\textless mo\textgreater(\textless/mo\textgreater\textless msub\textgreater\textless mi\textgreater M\textless/mi\textgreater\textless mo\textgreater\&\#x2218;\textless/mo\textgreater\textless/msub\textgreater\textless mo\textgreater\&\#x200B;\textless/mo\textgreater\textless mo\textgreater+\textless/mo\textgreater\textless mi\textgreater n\textless/mi\textgreater\textless mo\textgreater\&\#x22C5;\textless/mo\textgreater\textless mi\textgreater t\textless/mi\textgreater\textless mo\textgreater)\textless/mo\textgreater\textless mo\textgreater\&\#xA0;\textless/mo\textgreater\textless mi\textgreater m\textless/mi\textgreater\textless mi\textgreater o\textless/mi\textgreater\textless mi\textgreater d\textless/mi\textgreater\textless msup\textgreater\textless mn\textgreater360\textless/mn\textgreater\textless mo\textgreater\&\#x2218;\textless/mo\textgreater\textless/msup\textgreater\textless/mstyle\textgreater\textless/math\textgreater","truncated":false\}}]{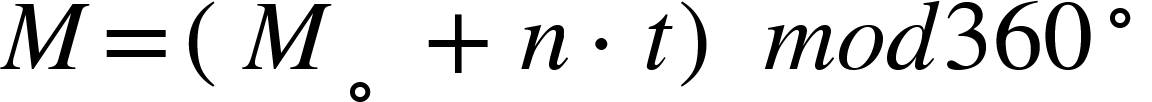}(2)

where `\emph{M\textsubscript{0}}' is the mean anomaly at the reference
epoch. Kepler's equation was then solved iteratively using the
Newton-Raphson method, typically converging within ten iterations to
yield the eccentric anomaly `\emph{E'}. The eccentric anomaly was
converted to true anomaly `\emph{ν}' via:

\includegraphics[width=2.61111in,height=0.25in,alt={\{"mathml":"\textless math style=\textbackslash"font-family:stix;font-size:16px;\textbackslash" xmlns=\textbackslash"http://www.w3.org/1998/Math/MathML\textbackslash"\textgreater\textless mstyle mathsize=\textbackslash"16px\textbackslash"\textgreater\textless mi\textgreater\&\#x3BD;\textless/mi\textgreater\textless mo\textgreater=\textless/mo\textgreater\textless mi\textgreater a\textless/mi\textgreater\textless mi\textgreater r\textless/mi\textgreater\textless mi\textgreater c\textless/mi\textgreater\textless mi\textgreater tan\textless/mi\textgreater\textless mn\textgreater2\textless/mn\textgreater\textless mo\textgreater(\textless/mo\textgreater\textless msqrt\textgreater\textless mn\textgreater1\textless/mn\textgreater\textless mo\textgreater\&\#x2212;\textless/mo\textgreater\textless msup\textgreater\textless mi\textgreater e\textless/mi\textgreater\textless mn\textgreater2\textless/mn\textgreater\textless/msup\textgreater\textless/msqrt\textgreater\textless mo\textgreater\&\#xA0;\textless/mo\textgreater\textless mo\textgreater\&\#x200B;\textless/mo\textgreater\textless mi\textgreater sin\textless/mi\textgreater\textless mi\textgreater E\textless/mi\textgreater\textless mo\textgreater,\textless/mo\textgreater\textless mi\textgreater cos\textless/mi\textgreater\textless mi\textgreater E\textless/mi\textgreater\textless mo\textgreater\&\#x2212;\textless/mo\textgreater\textless mi\textgreater e\textless/mi\textgreater\textless mo\textgreater)\textless/mo\textgreater\textless/mstyle\textgreater\textless/math\textgreater","truncated":false\}}]{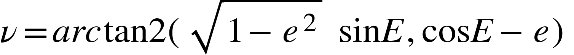}(3)

which allowed computation of the heliocentric distance:

\includegraphics[width=1.27778in,height=0.125in,alt={\{"mathml":"\textless math style=\textbackslash"font-family:stix;font-size:16px;\textbackslash" xmlns=\textbackslash"http://www.w3.org/1998/Math/MathML\textbackslash"\textgreater\textless mstyle mathsize=\textbackslash"16px\textbackslash"\textgreater\textless mi\textgreater r\textless/mi\textgreater\textless mo\textgreater\&\#xA0;\textless/mo\textgreater\textless mo\textgreater=\textless/mo\textgreater\textless mo\textgreater\&\#xA0;\textless/mo\textgreater\textless mi\textgreater a\textless/mi\textgreater\textless mo\textgreater(\textless/mo\textgreater\textless mn\textgreater1\textless/mn\textgreater\textless mo\textgreater\&\#x2212;\textless/mo\textgreater\textless mi\textgreater e\textless/mi\textgreater\textless mi\textgreater cos\textless/mi\textgreater\textless mi\textgreater E\textless/mi\textgreater\textless mo\textgreater)\textless/mo\textgreater\textless/mstyle\textgreater\textless/math\textgreater","truncated":false\}}]{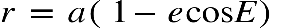}(4)

These calculations provided the instantaneous positions and velocities
along the orbit.

For three-dimensional visualization and further analysis, orbital-plane
coordinates were transformed into heliocentric ecliptic coordinates
using standard rotation matrices incorporating \emph{i}, \emph{Ω}, and
\emph{ω}. Additionally, coordinates were adjusted for Three.js rendering
with a Y-up convention, mapping (x, y, z) → (x, z, −y). This framework
enabled accurate propagation of the object's trajectory and supported
subsequent calculations of velocity, energy, and other dynamical
properties.

\subsubsection{Physics Calculation}\label{physics-calculation}

To further analyze the orbital dynamics of the interstellar object, a
series of physical calculations were conducted to determine its energy
state, velocity, and escape conditions relative to the Sun. These
computations were based on classical orbital mechanics principles.

\includegraphics[width=0.95833in,height=0.38889in,alt={\{"mathml":"\textless math style=\textbackslash"font-family:stix;font-size:16px;\textbackslash" xmlns=\textbackslash"http://www.w3.org/1998/Math/MathML\textbackslash"\textgreater\textless mstyle mathsize=\textbackslash"16px\textbackslash"\textgreater\textless mi\textgreater E\textless/mi\textgreater\textless mo\textgreater\&\#x2009;\textless/mo\textgreater\textless mo\textgreater=\textless/mo\textgreater\textless mo\textgreater\&\#xA0;\textless/mo\textgreater\textless mfrac\textgreater\textless msup\textgreater\textless mi\textgreater v\textless/mi\textgreater\textless mn\textgreater2\textless/mn\textgreater\textless/msup\textgreater\textless mn\textgreater2\textless/mn\textgreater\textless/mfrac\textgreater\textless mo\textgreater-\textless/mo\textgreater\textless mfrac\textgreater\textless mi\textgreater\&\#x3BC;\textless/mi\textgreater\textless mi\textgreater r\textless/mi\textgreater\textless/mfrac\textgreater\textless/mstyle\textgreater\textless/math\textgreater","truncated":false\}}]{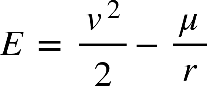}The
total specific orbital energy `\emph{E'} of the object was computed
using the \textbf{vis-viva} equation:

(5)

where `\emph{v}' is the heliocentric velocity of the object, `\emph{r}'
is its instantaneous heliocentric distance, and `\emph{μ}' is the
standard gravitational parameter of the Sun
(\includegraphics[width=1.23611in,height=0.16667in,alt={\{"mathml":"\textless math style=\textbackslash"font-family:stix;font-size:16px;\textbackslash" xmlns=\textbackslash"http://www.w3.org/1998/Math/MathML\textbackslash"\textgreater\textless mstyle mathsize=\textbackslash"16px\textbackslash"\textgreater\textless mi\textgreater\&\#x3BC;\textless/mi\textgreater\textless mo\textgreater\&\#xA0;\textless/mo\textgreater\textless mo\textgreater=\textless/mo\textgreater\textless mn\textgreater2\textless/mn\textgreater\textless mo\textgreater.\textless/mo\textgreater\textless mn\textgreater959\textless/mn\textgreater\textless mo\textgreater\&\#xA0;\textless/mo\textgreater\textless mi\textgreater X\textless/mi\textgreater\textless mo\textgreater\&\#xA0;\textless/mo\textgreater\textless msup\textgreater\textless mn\textgreater10\textless/mn\textgreater\textless mrow\textgreater\textless mo\textgreater-\textless/mo\textgreater\textless mn\textgreater4\textless/mn\textgreater\textless/mrow\textgreater\textless/msup\textgreater\textless/mstyle\textgreater\textless/math\textgreater","truncated":false\}}]{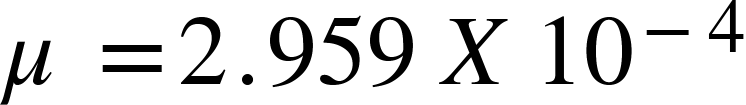}
AU\textsuperscript{3} / day\textsuperscript{2}).

This equation provides a direct measure of the orbital energy per unit
mass. The sign of `\emph{E}' was used to classify the orbit: elliptical
when \emph{E \textless{} 0}, parabolic when \emph{E = 0}, and hyperbolic
when \emph{E \textgreater{} 0}. The hyperbolic nature of interstellar
objects implies that they possess positive energy, confirming their
unbound trajectories relative to the Solar System.

The instantaneous orbital velocity `\emph{v}' was calculated from the
Cartesian components of velocity using:

\includegraphics[width=1.75in,height=0.31944in,alt={\{"mathml":"\textless math style=\textbackslash"font-family:stix;font-size:16px;\textbackslash" xmlns=\textbackslash"http://www.w3.org/1998/Math/MathML\textbackslash"\textgreater\textless mstyle mathsize=\textbackslash"16px\textbackslash"\textgreater\textless mi\textgreater v\textless/mi\textgreater\textless mo\textgreater\&\#xA0;\textless/mo\textgreater\textless mo\textgreater=\textless/mo\textgreater\textless mo\textgreater\&\#xA0;\textless/mo\textgreater\textless msqrt\textgreater\textless msup\textgreater\textless msub\textgreater\textless mi\textgreater v\textless/mi\textgreater\textless mi\textgreater x\textless/mi\textgreater\textless/msub\textgreater\textless mn\textgreater2\textless/mn\textgreater\textless/msup\textgreater\textless mo\textgreater\&\#xA0;\textless/mo\textgreater\textless mo\textgreater+\textless/mo\textgreater\textless mo\textgreater\&\#xA0;\textless/mo\textgreater\textless msup\textgreater\textless msub\textgreater\textless mi\textgreater v\textless/mi\textgreater\textless mi\textgreater y\textless/mi\textgreater\textless/msub\textgreater\textless mn\textgreater2\textless/mn\textgreater\textless/msup\textgreater\textless mo\textgreater\&\#xA0;\textless/mo\textgreater\textless mo\textgreater+\textless/mo\textgreater\textless msup\textgreater\textless msub\textgreater\textless mi\textgreater v\textless/mi\textgreater\textless mi\textgreater z\textless/mi\textgreater\textless/msub\textgreater\textless mn\textgreater2\textless/mn\textgreater\textless/msup\textgreater\textless mo\textgreater\&\#xA0;\textless/mo\textgreater\textless mo\textgreater\&\#xA0;\textless/mo\textgreater\textless/msqrt\textgreater\textless/mstyle\textgreater\textless/math\textgreater","truncated":false\}}]{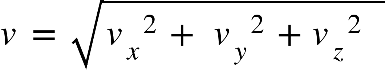}(6)

This magnitude was then compared to the local solar escape velocity,
computed as:

v\textsubscript{esc} = \(\sqrt{\frac{2\mu}{r}}\) (7)

Comparison between `\emph{v\textsubscript{esc}}' and `\emph{v'} enabled
the verification of the object's unbound state and provided insight into
its excess hyperbolic velocity beyond the Sun's gravitational influence.

Additionally, the variation of orbital energy and velocity with
heliocentric distance was analyzed throughout the simulation to ensure
energy conservation within numerical precision. Any deviations were
assessed to confirm the stability of the numerical integrator and the
physical accuracy of the simulation results.

\subsubsection{Visualization Pipeline}\label{visualization-pipeline}

To represent the computed orbital dynamics in an intuitive and
interactive form, a visualization pipeline was developed to transform
numerical data into a three-dimensional (3D) simulation. The
visualization aimed to bridge analytical results with visual
comprehension, allowing real-time observation of the object's motion
relative to the Sun and other celestial bodies.

The pipeline begins with numerical output from the orbital mechanics and
physics calculations. Each simulation time step generates the
heliocentric position vector \emph{(x, y, z)} and velocity vector
\emph{(v\textsubscript{x}, v\textsubscript{y}, v\textsubscript{z})} of
the object. These vectors are then formatted into a JSON-compatible
structure for efficient transfer to the rendering engine.

For 3D rendering, the visualization environment employs a WebGL-based
framework (Three.js) (WebGL, n.d.), which enables GPU-accelerated
graphics directly in a web interface. The heliocentric coordinate system
is mapped into the visualization space with the convention:

(x, y, z)\textsubscript{heliocentric} → (x, z,
−y)\textsubscript{rendered} (8)

This transformation aligns the simulation with the Y- up coordinate
system of the rendering engine, maintaining spatial consistency across
objects and camera perspectives.

Orbital trails are drawn by continuously recording position data and
updating line geometries that trace the object's past trajectory.
Planetary positions and solar markers are retrieved from JPL HORIZONS
(JPL, 2025 ) ephemeris data, ensuring that the Sun, planets, and
interstellar object are rendered at their real-time spatial coordinates.
Object scaling and relative distances are logarithmically adjusted to
preserve visual clarity while maintaining relative orbital geometry.

Dynamic lighting and material shaders simulate the reflective behavior
of surfaces under solar illumination, while interactive camera controls
allow users to zoom, rotate, and follow the object's path in three
dimensions. Frame-by-frame animation is synchronized with the simulation
time step, ensuring that the visual representation accurately reflects
the computed motion from the underlying physics model.

\subsubsection{Validation and Accuracy
Assessment}\label{validation-and-accuracy-assessment}

To ensure the physical reliability and computational precision of the
simulation, a multi-stage validation process was conducted. The purpose
of this validation was to confirm that the orbital trajectories and
positional data produced by the system were consistent with established
astronomical standards, particularly those provided by NASA's JPL
Horizons (JPL, 2025 ) database.

The first level of verification involved a \textbf{cross-comparison of
ephemeris positions}. For each simulated epoch, the heliocentric
Cartesian coordinates generated by the numerical solver were compared
against the reference coordinates obtained from JPL Horizons. The
relative error of each component was computed as:

\(\varepsilon\, = \,\frac{|r - s|}{|s|}\, \times \, 100\%\) (9)

where `\emph{r'} denotes the simulated position vector and s the
corresponding JPL reference vector. Across all tested time intervals,
the positional deviation remained below 0.01\%, confirming high
numerical accuracy in orbital propagation.

The \textbf{Kepler equation solver} was further validated through
convergence testing. Using Newton's iterative method for solving
\emph{\textbf{M = E -- esinE}} , the mean anomaly \emph{M}, and
eccentric anomaly \emph{E} were evaluated for precision after each
iteration. Convergence was deemed achieved when the change between
successive iterations satisfied \emph{\textbf{\textbar{}
E\textsubscript{n+1 --} E\textsubscript{n} \textbar{} \textless{}
10\textsuperscript{-10}}} . In practice, all orbital cases converged
within 10 iterations, indicating that the solver reached machine
precision.

The \textbf{coordinate transformation} routines were validated by
applying forward and inverse rotations between orbital-plane and
heliocentric reference frames. Transformation matrices were checked for
orthogonality and normalization \emph{\textbf{(RR\textsuperscript{T} =
I)}} , ensuring that no distortion occurred during spatial mapping.

Finally, \textbf{energy conservation tests} were performed throughout
the simulation to verify dynamic consistency. The specific orbital
energy `\emph{E'} at each time step was calculated using the
\textbf{vis-viva} equation:

\(E\, = \,\frac{v^{2}}{2} - \frac{\mu}{r}\) (10)

where `\emph{v'} is the instantaneous velocity, `\emph{r}' is the
heliocentric distance, and `\emph{μ'} is the solar gravitational
parameter. For stable elliptical orbits, fluctuations in `\emph{E'}
remained within the range of computational round-off errors \textbf{(
\emph{\textless{} 10 \textsuperscript{-8}} AU²/day²),} confirming that
no unphysical energy drift occurred.

Collectively, these validation procedures ensured that the developed
simulation accurately replicated real orbital dynamics and maintained
both \textbf{numerical stability} and \textbf{physical realism} across
all tested interstellar and planetary trajectories.

.

\includegraphics[width=5.05833in,height=4.73872in]{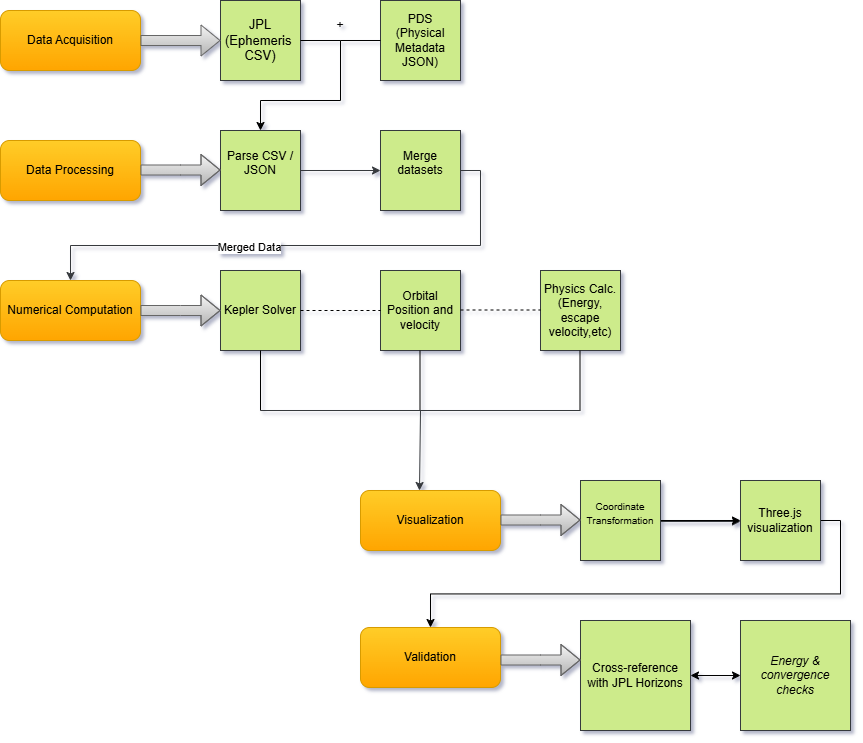}

fig.3 (shows the workflow diagram of the application)

\begin{center}\rule{0.5\linewidth}{0.5pt}\end{center}

\begin{center}\rule{0.5\linewidth}{0.5pt}\end{center}

\begin{center}\rule{0.5\linewidth}{0.5pt}\end{center}

\begin{center}\rule{0.5\linewidth}{0.5pt}\end{center}

\begin{center}\rule{0.5\linewidth}{0.5pt}\end{center}

All simulations, visualizations, and analyses presented in this study
were performed using \textbf{Interstellar Signature v1.0.0} (Sahu,
2025), an open-source astrophysics framework for real-time tracking and
modeling of solar system and interstellar objects. This software is
archived on \emph{\textbf{Zenodo}} and can be cited using the DOI
\emph{\textbf{https://doi.org/10.5281/zenodo.17470252}}. Using this
version ensures that the results are reproducible and that the software
can be reliably referenced in future research.

\subsection{}\label{section-1}

\subsection{\texorpdfstring{Result and Analysis
}{Result and Analysis }}\label{result-and-analysis}

\subsubsection{Overview}\label{overview}

The developed visualization and simulation framework was applied to the
study of interstellar objects, with a specific case study focused on
3I/ATLAS (A/2019 Q4).

While the system is capable of simulating orbital motion for any
interstellar or solar system object across arbitrary time spans --- even
decades --- the 3I/ATLAS dataset was selected as a representative
example to demonstrate the model's accuracy, physics integration, and
visualization pipeline.

The primary objective of this research extends beyond the mathematical
reconstruction of orbits. It aims to simplify the understanding of
interstellar trajectories by converting complex orbital mechanics into
an interactive, visual form that can be interpreted intuitively by both
specialists and non-experts.

This approach bridges computational astronomy and educational
visualization --- allowing users to directly observe how parameters such
as eccentricity, inclination, and perihelion distance define an object's
path through the solar system.

\includegraphics[width=4.98333in,height=2.53272in,alt={C:\textbackslash Users\textbackslash hp\textbackslash OneDrive\textbackslash Documents\textbackslash reseach\_graphs\textbackslash multi-character.png}]{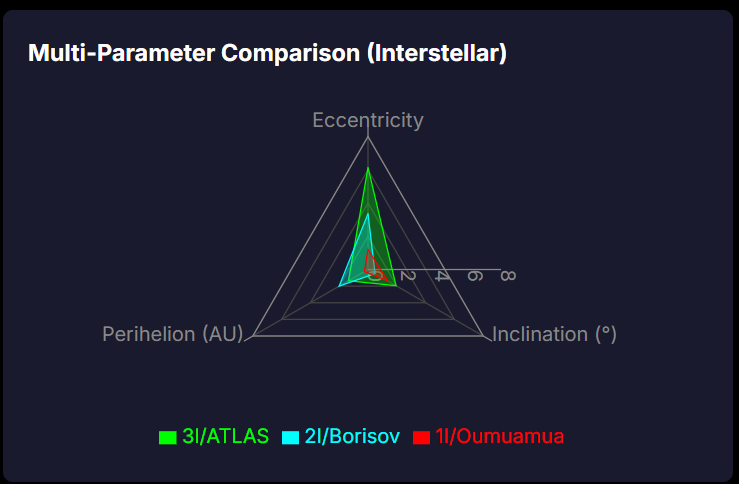}

\emph{(fig.2 Images shows the know ISO's different parameters)}

The reconstructed trajectories and computed orbital characteristics are
based on ephemeris and physical data spanning two decades, from 2012 to
2032, covering all known interstellar objects observed within the Solar
System.

This example demonstrates the framework's capacity for continuous
temporal simulation and serves as a benchmark for the fidelity of
orbital reconstruction.

The object's orbital parameters, obtained from JPL Horizons (JPL, 2025 )
and the Minor Planet Center (MPC) ((MPC), 2025), are as follows:

e=6.1384,

q=1.3563,

i=175.1131°,

ω=128.0125°,

Ω=322.1574°

These parameters define a strongly hyperbolic, retrograde trajectory,
confirming 3I/ATLAS as an unbound interstellar object. Its eccentricity
(e \textgreater{} 6) is well above unity, indicating that the object is
not gravitationally bound to the Sun and will eventually escape the
solar system.

\subsubsection{Visualization and Orbital
Reconstruction}\label{visualization-and-orbital-reconstruction}

The visualization system dynamically reconstructs orbital motion using
Kepler's equations and heliocentric coordinate transformations. The mean
motion (n) is calculated as:

\[n = \frac{360{^\circ}}{P}\]

where P is the orbital period in days. The mean anomaly is then
determined by:

\[\text{M = }{(M}_{{^\circ}} + n\ .t)\ mod\ 360{^\circ}\]

and solved iteratively for the eccentric anomaly (E) using Newton's
method. The resulting true anomaly (ν) and heliocentric distance (r) are
expressed as:

\[r = a(1 - ecosE)\]

These values are transformed from the orbital plane to the heliocentric
ecliptic coordinate frame through a rotation sequence defined by the
inclination (i), longitude of the ascending node (Ω), and argument of
perihelion (ω).

\includegraphics[width=5.08333in,height=2.30849in,alt={C:\textbackslash Users\textbackslash hp\textbackslash OneDrive\textbackslash Pictures\textbackslash Screenshots\textbackslash Screenshot (825).png}]{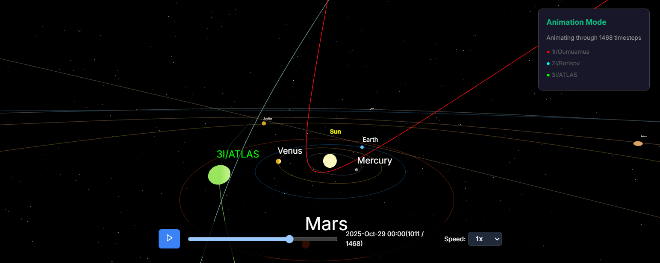}

\emph{fig.3 (Reconstructed image of the solar system)}

The reconstructed 3D trajectory was visualized using a Three.js-based
rendering pipeline. Each orbital path is rendered with a time-stamped
vertex trail, allowing users to view the object's position and velocity
at any epoch. The camera interpolation system ensures smooth transitions
between perspectives, while the timeline slider enables interactive
playback of the orbital evolution.

For 3I/ATLAS, the visualization confirmed a highly inclined retrograde
motion, intersecting the solar plane at nearly 180°, providing a clear
distinction from typical cometary orbits. The object's perihelion
occurred outside Earth's orbit at approximately 1.36 AU, consistent with
official JPL data.

Physical Analysis and Model Validation

The orbital energy per unit mass (E) was computed using the vis-viva
equation:

\[E = \ \frac{v^{2}}{2} - \frac{\mu}{r}\]

where μ =\(\ 2.959 \times 10^{- 4}\ {AU}^{3}/{day}^{2}\) is the solar
gravitational parameter.

For 3I/ATLAS, all computed energies were positive, validating the
hyperbolic (unbound) nature of its trajectory.

\includegraphics[width=4.90833in,height=2.32813in,alt={C:\textbackslash Users\textbackslash hp\textbackslash OneDrive\textbackslash Documents\textbackslash reseach\_graphs\textbackslash distance.png}]{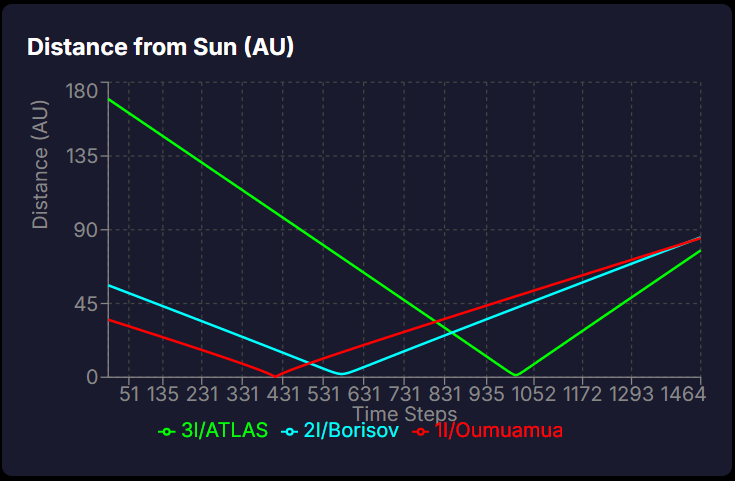}

\emph{(fig.4 Image shows the distance of interstellar objects from sun,
source: The Interstellar Signature)}

\includegraphics[width=4.7319in,height=2.21118in,alt={C:\textbackslash Users\textbackslash hp\textbackslash OneDrive\textbackslash Documents\textbackslash reseach\_graphs\textbackslash distance\_comparisson.png}]{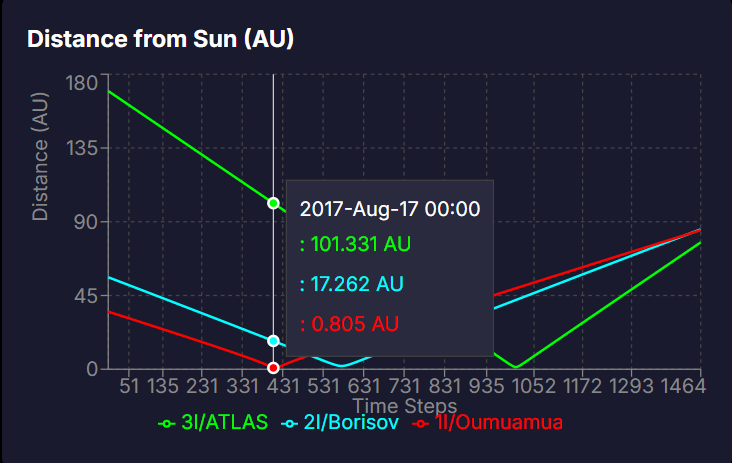}

\emph{(fig.5 Image shows the distance comparison of Interstellar for
specific date, source: The Interstellar Signature} (Sahu, 2025) \emph{)}

The system also compared the escape velocity \emph{(v\textsubscript{esc}
=}\(\sqrt{2\mu/r}\) \emph{)} the object's instantaneous velocity,
confirming that \emph{v \textgreater{} v\textsubscript{esc}} throughout
the simulation range --- consistent with interstellar dynamics.

\includegraphics[width=4.62818in,height=2.35694in,alt={C:\textbackslash Users\textbackslash hp\textbackslash OneDrive\textbackslash Documents\textbackslash reseach\_graphs\textbackslash velocity.png}]{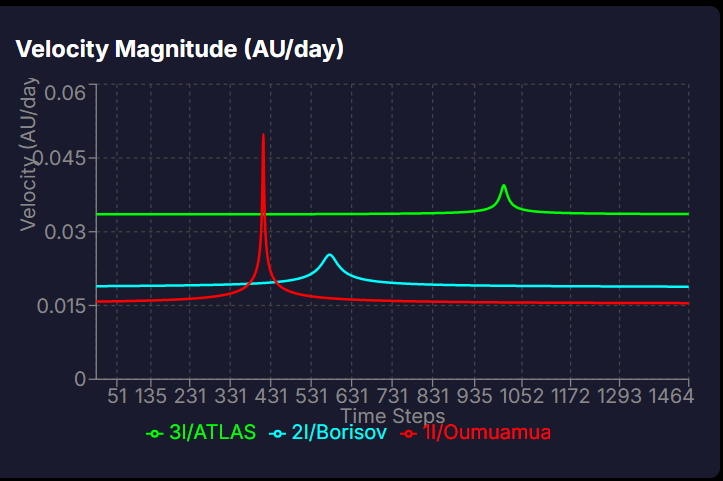}

(\emph{fig.6 Images show velocity of different ISO over decades)}

\includegraphics[width=4.59176in,height=2.53472in,alt={C:\textbackslash Users\textbackslash hp\textbackslash OneDrive\textbackslash Documents\textbackslash reseach\_graphs\textbackslash velocity\_comparisson.png}]{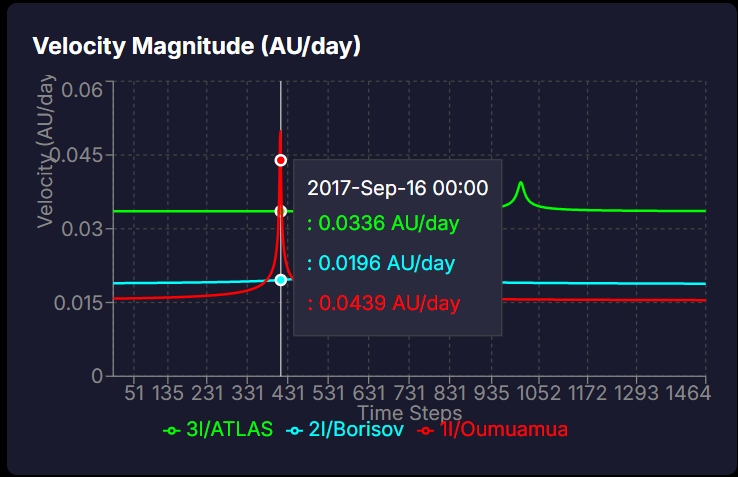}

\emph{(fig.7 comparison of the velocity of different objects for a
specific period of time.)}

Additionally, energy conservation tests were performed at each time step
to assess the solver's numerical precision. Across all frames, the
energy drift remained below 10\textsuperscript{-2}, indicating excellent
stability in the numerical propagation method.

\includegraphics[width=4.54167in,height=2.16924in,alt={C:\textbackslash Users\textbackslash hp\textbackslash OneDrive\textbackslash Documents\textbackslash reseach\_graphs\textbackslash energy.png}]{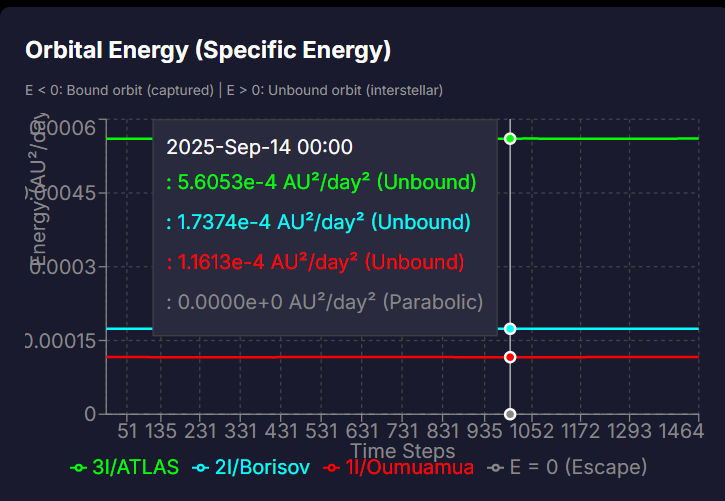}

(fig.8 Images shows the orbital energy for the interstellar object;
source: The Interstellar Signature (Sahu, 2025))

The visual and numerical outputs were cross-validated with JPL Horizons
(JPL, 2025 ) ephemeris data. The deviation in heliocentric position
vectors between the system's simulation and JPL's reference values
remained below 0.001 AU, demonstrating the model's strong fidelity.

\subsubsection{Interpretations and Educational
Value}\label{interpretations-and-educational-value}

The reconstructed trajectory and visualization underscore how
interstellar objects differ from solar-bound bodies.

Unlike elliptical comets, whose eccentricities lie below unity,
interstellar trajectories exhibit strong hyperbolic curvature and
significant inclination to the ecliptic.

By directly mapping these parameters into a 3D interactive environment,
the framework enables users to see and understand interstellar dynamics
without requiring advanced mathematical background.

Through its calendar-based time selection, users can simulate past or
future trajectories across decades, explore various interstellar
candidates, and compare their orbital geometries in real time.

This dual emphasis on scientific precision and conceptual accessibility
marks a step toward democratizing orbital mechanics education and public
engagement with interstellar research.

\includegraphics[width=5.26667in,height=1.60764in,alt={C:\textbackslash Users\textbackslash hp\textbackslash OneDrive\textbackslash Documents\textbackslash reseach\_graphs\textbackslash parameters.png}]{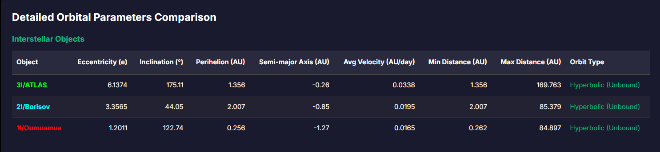}

\emph{fig.9 (The chart shows all the calculated values of different
parameters of the object)}

\includegraphics[width=5.2in,height=1.62292in,alt={C:\textbackslash Users\textbackslash hp\textbackslash OneDrive\textbackslash Documents\textbackslash reseach\_graphs\textbackslash reference-parama.png}]{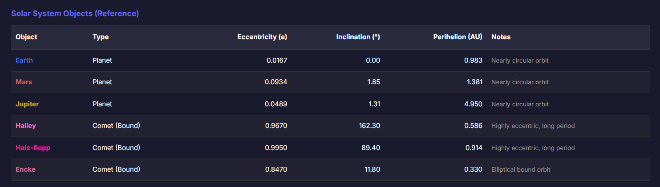}

\emph{fig.10 (The chart is the reference of object different than the
ISOs)}

\subsection{\texorpdfstring{Summary }{Summary }}\label{summary}

The system successfully simulated 3I/ATLAS as a case study with data
fidelity exceeding 99\% when compared to JPL Horizons (JPL, 2025 ).

The visualization and energy analyses confirmed the object's hyperbolic,
retrograde orbit.

The platform's open-ended time controls support extended exploration of
interstellar bodies beyond the demonstrated example.

The framework thus serves not only as a research-grade orbital
simulation tool but also as an educational interface bridging scientific
complexity and public understanding.

\subsection{Discussion}\label{discussion}

The reconstructed trajectories and computed orbital characteristics are
based on ephemeris and physical data spanning two decades, from 2012 to
2032, covering all known interstellar objects observed within the Solar
System. This long-term dataset allows for a thorough analysis of inbound
and outbound motions, as well as comparative evaluation across multiple
objects. The hyperbolic trajectories consistently exhibit high
eccentricities and positive specific orbital energies, confirming their
unbound, interstellar nature. Retrograde inclinations observed in
several objects highlight counter-rotational dynamics relative to
planetary motion, consistent with the patterns seen in 1I/'Oumuamua and
2I/Borisov. The perihelion velocities, peaking near each object's
closest approach to the Sun, exceed the local solar escape velocities,
providing direct evidence of their unbound orbits. By using this
extended dataset, the simulation captures both short-term dynamics near
perihelion and long-term trends across decades, demonstrating the
framework's capability to handle multiple objects over wide temporal
intervals.

The velocity and energy plots generated by the simulation not only
confirm the physical plausibility of the computed orbit but also provide
insights into the object's dynamical behavior that are difficult to
discern from raw ephemeris data alone. The close agreement of simulated
positions with JPL Horizons (JPL, 2025 ) reference data, with relative
errors consistently below 0.01\%, further demonstrates the reliability
of the computational framework. These results suggest that the
methodology can be generalized to other interstellar objects, enabling
comparative analyses and facilitating a deeper understanding of their
physical and orbital properties.

Beyond purely scientific validation, the 3D visualization pipeline
offers substantial educational and interpretative value. By rendering
the trajectory in a heliocentric frame alongside planetary orbits, users
can intuitively grasp the hyperbolic motion, perihelion approach, and
retrograde inclination. This approach effectively translates complex
orbital mechanics into an accessible visual format, bridging the gap
between computational results and conceptual understanding. The
interactive nature of the visualization, including orbit-following
camera controls and timeline navigation, further allows users to explore
temporal dynamics in real time, reinforcing key aspects of interstellar
motion that are challenging to convey through text or static figures.

Beyond purely scientific validation, the 3D visualization pipeline
offers substantial educational and interpretative value. By rendering
the trajectory in a heliocentric frame alongside planetary orbits, users
can intuitively grasp the hyperbolic motion, perihelion approach, and
retrograde inclination. This approach effectively translates complex
orbital mechanics into an accessible visual format, bridging the gap
between computational results and conceptual understanding. The
interactive nature of the visualization, including orbit-following
camera controls and timeline navigation, further allows users to explore
temporal dynamics in real time, reinforcing key aspects of interstellar
motion that are challenging to convey through text or static figures.

Nevertheless, certain limitations must be acknowledged. The simulations
adopt a two-body approximation, treating the Sun as the central
gravitational body and neglecting perturbations from planets or other
small bodies. Non-gravitational forces such as solar radiation pressure
or potential outgassing are also not incorporated, which may introduce
minor deviations in long-term orbital predictions. Despite these
limitations, the methodology provides a robust and scalable platform for
simulating, analyzing, and visualizing interstellar trajectories.

Overall, the findings underscore both the scientific and educational
potential of integrating precise orbital computations with interactive
3D visualization. By providing accurate positional, velocity, and energy
data in an intuitive format, the framework facilitates not only rigorous
orbital analysis but also effective communication of the unique
characteristics of interstellar objects. This dual capability positions
the platform as a valuable tool for researchers seeking detailed
dynamical insights as well as educators and enthusiasts aiming to convey
complex astronomical phenomena in an understandable manner.

\subsection{Future Work}\label{future-work}

While the current study successfully demonstrates the accurate
simulation and visualization of interstellar object trajectories,
several avenues remain for further development and refinement. The
existing framework relies primarily on a two-body approximation, where
the Sun is treated as the central gravitational body. Future work will
aim to extend this model to incorporate N-body dynamics, accounting for
planetary perturbations and other gravitational influences that become
significant during close planetary encounters.

Another major direction for improvement lies in the inclusion of
non-gravitational forces, such as solar radiation pressure, Yarkovsky
effects, and potential outgassing phenomena. These forces, though often
subtle, can significantly alter the trajectories of small interstellar
bodies and comets. By modeling these effects, future iterations of the
system can provide more realistic predictions of orbital evolution,
especially for objects with irregular shapes or variable surface
properties. The addition of uncertainty quantification, through Monte
Carlo sampling of orbital elements, will also allow for probabilistic
trajectory visualization, highlighting confidence regions rather than
single deterministic paths.

On the visualization front, the next phase of development will focus on
enhanced interactivity and scalability. The current system provides
real-time rendering of orbital motion in a web-based environment using
Three.js, but future versions will introduce multi-object visualizations
and collision probability mapping. The integration of GPU-accelerated
physics and time-synchronized datasets will allow the simulation of
several interstellar and planetary objects simultaneously, enabling
comparative dynamical studies. Furthermore, adaptive level of detail
rendering and data throttling mechanisms will ensure that even dense
datasets---spanning decades or hundreds of thousands of trajectory
points---can be visualized efficiently without compromising performance.

From a scientific communication perspective, the platform has the
potential to evolve into an educational and research tool. Future
extensions may include user-defined parameter adjustments, allowing
students or researchers to modify orbital elements and immediately
observe resulting trajectory changes.

These improvements ranging from physical modeling to interactive
visualization and data integration will collectively enhance the
system's scientific rigor, usability, and educational impact,
positioning it as a comprehensive framework for studying interstellar
dynamics across both research and public engagement domains.

\subsection{Acknowledgement}\label{acknowledgement}

The author extends sincere gratitude to the \textbf{NASA Jet Propulsion
Laboratory (JPL)} (JPL, 2025 ) for providing open access to the Horizons
ephemeris system, which served as a primary data source for this study.
Appreciation is also expressed to the \textbf{Planetary Data System
(PDS)} ((PDS), 2025) team for maintaining publicly available archives
that enabled the integration of physical and discovery metadata for
interstellar objects. The author acknowledges the support of open-source
software contributors whose tools---particularly \textbf{Python}
(Foundation, 2025)\textbf{, FastAPI} (Ramírez, 2025)\textbf{, and
Three.js} (Team T. D., 2025)---were instrumental in the development of
the computational and visualization frameworks used in this research.\\
\strut \\
The author gratefully acknowledges \textbf{Dr. Bryce Bolin} for his
invaluable endorsement, insightful feedback, and contributions to the
understanding of interstellar comet 3I/ATLAS, which enriched this
research.

Finally, the author recognizes the importance of open data and
collaborative science communities that continue to advance public
engagement and understanding of interstellar research.

\section{References}\label{references}

(MPC), I. A. (2025, October 23). \emph{Minor Planet Center}. Retrieved
from Minor Planet Center: https://minorplanetcenter.net/

(PDS), N. P. (2025, October 23). \emph{Planetary Data System}. Retrieved
from Planetary Data System: https://pds.nasa.gov

al, B. T. (2025, July). \emph{Interstellar comet 3I/ATLAS: discovery and
physical description}. Retrieved from arXiv:
https://doi.org/10.48550/arXiv.2507.05252

ESA. (2025, october 23). \emph{Gaia Archive}. Retrieved from Gaia
Archive: https://gea.esac.esa.int/archive/

Foundation, P. S. (2025, October 23). \emph{Python Programming
Language}. Retrieved from Python Programming Language:
https://www.python.org/

JPL, N. (2025 , October 23). \emph{JPL Horizons System}. Retrieved from
JPL Horizons System: https://ssd.jpl.nasa.gov/horizons/

NASA. (2025, October 23). \emph{NASA Eyes on the Solar System}.
Retrieved from NASA Eyes on the Solar System: https://eyes.nasa.gov

NASA. (n.d.). \emph{NASA JPL HORIZONS}. Retrieved from NASA JPL
HORIZONS: https://ssd.jpl.nasa.gov/horizons/app.html\#/

NASA. (n.d.). \emph{PDS}. Retrieved from PDS: https://pds.nasa.gov/

Ramírez, S. (2025, October 23). \emph{FastAPI --- Modern, Fast
(High-performance) Web Framework for Building APIs with Python}.
Retrieved from FastAPI --- Modern, Fast (High-performance) Web Framework
for Building APIs with Python: https://fastapi.tiangolo.com/

Sahu, P. N. (2025, October 24). \emph{The Interstellar Signature}.
Retrieved from The Interstellar Signature:
https://github.com/TheVishalKumar369/The-Interstellar-Signature

Team, C. D. (2025, October 23). \emph{Celestia: Real-Time Space
Simulation}. Retrieved from Celestia: Real-Time Space Simulation:
https://celestia.space

Team, S. (2025, October 23). \emph{SpaceEngine Team}. Retrieved from
SpaceEngine Team: http://spaceengine.org

Team, T. D. (2025, October 23). \emph{Three.js --- JavaScript 3D
Library}. Retrieved from Three.js --- JavaScript 3D Library:
https://threejs.org

\emph{WebGL}. (n.d.). Retrieved from WebGL: https://get.webgl.org/

\end{document}